\documentclass[prb]{revtex4}% Physical Review B
\usepackage{amsfonts,natbib}
\usepackage{amsmath}
\usepackage{amssymb}
\usepackage{graphicx}
\usepackage{amsfonts}%
\renewcommand{\vec}[1]{\boldsymbol{\mathbf{#1}}}

\begin{document}

\title[Short title for running header]{Quantum Transport in a Nanosize Silicon-on-Insulator Metal-Oxide-Semiconductor
Field Effect Transistor}
\author{M. D. Croitoru$^{\dag}$, V. N. Gladilin$^{\dag}$, V. M. Fomin$^{\dag}$, J. T.
Devreese$^{\ddag}$}
\affiliation{Theoretische Fysica van de Vaste Stoffen (TFVS), Universiteit Antwerpen (UIA),
Universiteitsplein 1, B-2610 Antwerpen, Belgium}
\author{W. Magnus, W. Schoenmaker, and B. Sor\'{e}e}
\affiliation{IMEC, B-3001 Leuven,~Belgium}

\begin{abstract}
An approach is developed for the determination of the current flowing through
a nanosize silicon-on-insulator (SOI) metal-oxide-semiconductor field-effect
transistors (MOSFET). The quantum mechanical features of the electron
transport are extracted from the numerical solution of the quantum Liouville
equation in the Wigner function representation. Accounting for electron
scattering due to ionized impurities, acoustic phonons and surface roughness
at the Si/SiO$_{2}$ interface, device characteristics are obtained as a
function of a channel length. From the Wigner function distributions, the
coexistence of the diffusive and the ballistic transport naturally emerges. It
is shown that the scattering mechanisms tend to reduce the ballistic component
of the transport. The ballistic component increases with decreasing the
channel length.

%Shell document for REV\TeX{} 4.

\end{abstract}
\volumeyear{year}
\volumenumber{number}
\issuenumber{number}
\eid{identifier}
\date{\today}
\startpage{1}
\endpage{2}
\maketitle

\section{Introduction}

Down scaling MOSFETs to their limiting sizes is a key challenge of the
semiconductor industry. Detailed simulations that capture the transport and
the quantum mechanical effects that occur in these devices can complement
experimental work in addressing this challenge. Furthermore, a conceptual view
of the nanoscale transistor is needed to support the interpretation of the
simulations and experimental data as well as to guide further experimental
work. The objective of this work is to provide such a view by formulating a
detailed quantum-mechanical transport model and performing extensive numerical
simulations. We develop a model along these lines for the nanosize SOI, or
thin film MOSFET, since the suppression of the short-channel effect seems to
be most adequately addressed by this technology. Therefore, understanding the
SOI MOSFET including its quantum-mechanical transport properties is urgently
needed. In this work, we restrict our attention to the steady-state
current-voltage characteristics.

The SOI MOSFET \cite{Sze} is a transistor, built on a thin silicon layer,
which is separated from the substrate by an insulating layer (Fig. 1). In the
nanoscale transistor, we have to optimize device functionality and
reliability. To achieve this goal, we need to suppress any short-channel
effects as much as possible. These effects include threshold voltage
variations versus channel length and the drain-induced barrier lowering
effect, i.e. a drain voltage dependence of threshold voltage, which
complicates transistor design at a circuit level. Frank \textit{et
al.}\cite{Frank} recently showed, that increasing the channel doping
concentration, $N_{\mathrm{CH}}$, suppresses short-channel effects. The
integration of the down-scaled devices requires that the gate insulating layer
thickness should be reduced and/or higher dielectric constant of the insulator
are implemented.

Si-based MOSFETs with typical sizes about 100~nm have found an application in
highly integrated systems. Mechanisms of the electron transport in these
devices differ from those in the devices of 50~nm and below. The
'conventional' devices are described by the Boltzmann transport equation and
approximation thereof. This theories focus on scattering-dominant transport,
which typically occurs in long channel devices. On the contrary, in a
structure with a characteristic size of the order of the mean free path, the
electron transport is essentially quasi-ballistic.

To describe the electron transport through a nanoscale transistor, a
quantum-mechanical treatment is required. This includes the following
approaches: the non-equilibrium Green's function formalism of Keldysh and
Kadanoff-Baym
\cite{Keldysh,Kadanoff,Han,Jauho1,Sarker,Jauho2,Datta1,Datta2,Ren,Lu}, the
Pauli master equation method \cite{Fischetti1,Fischetti2,Castella}, the full
density matrix method \cite{Brun,Jacob} and Wigner function method
\cite{Ferry,Frensley1,Kluk,Buot,Frensley,Tsu,old}.

The main goal of the present work is the investigation of quantum transport in
a nanosize MOSFET. Consequently, quantum effects such as quantum reflection
and quantum tunnelling, occur in the electron transport. Simultaneously,
electron scattering should be taken into account. Due to a low
electron-scattering rate, the sub-50~nm devices are expected to offer higher
speed performance than the conventional devices do. In a conventional MOSFET,
the conductance is determined by the scattering rate of electrons in the
channel, and the velocity of charge carriers is limited to a maximal value of
about 10$^{7}$~cm/s. In MOSFETs with the channel length less than the mean
free path, a velocity overshoot effect occurs \cite{Mizuno}. The small length
of the channel results into an increased on-state current and a reduced gate
capacitance. For example, in Ref.\thinspace\cite{Nakajima} the fabrication of
sub-0.1~$\mu$m MOSFETs was reported. The resulting devices demonstrate a
rather good performance in a wide range of temperatures.

In \cite{old}, a flexible two-dimensional model has been developed, that
optimally combines analytical and numerical methods and describes the key
features of cylindrical SOI MOSFET devices. The theoretical tools to describe
the quantum transport features involve the Wigner distribution function
formalism~\cite{Frensley}, in which a Boltzmann collision term representing
acoustics phonons, impurity and surface-roughness scattering was incorporated.
Therefore this model is capable to probe the competition between three
effects: quantum reflection, quantum tunnelling and phase de-coherence due to
elastic scattering. The Wigner function formalism offers many advantages for
quantum modelling \cite{who,Bordone}. First, it is a phase-space distribution,
similar to classical distributions. Because of the phase-space nature of the
distribution, it is conceptually possible to use the correspondence principle
in order to determine, where quantum corrections enter the problem. At the
boundaries, the phase-space description allows for separation of incoming and
outgoing components of the distribution, which therefore permits the modelling
an ideal contact, and hence an open system \cite{Kluk}. By coupling the
equation for the Wigner function to the Poisson equation, we obtain a
fully-self-consistent model of the SOI MOSFET.

This paper is organized as follows~: in Section~II, a description of the
system is presented in terms of a one-electron Hamiltonian. In Section~III the
quantum Liouville equation satisfied by the electron density matrix is
transformed into a set of one-dimensional equations for partial Wigner
functions. The boundary conditions for the Wigner function are considered in
Section IV. The ballistic regime of the structure described in Section V. A
one-dimensional collision term is derived in Section~VI, in which we extend
earlier work \cite{old} by also including scattering due to interface
roughness. In Section VII, we describe a numerical model to solve the
equations for the Wigner function. In Section~VIII, the results of the
numerical calculations are discussed.

\section{The Hamiltonian of the system}

We consider a $n$-channel SOI MOSFET structure. When a positive gate voltage
is applied, electrons in the channel are confined to a narrow inversion layer
near Si/SiO$_{2}$ interface. The current flows in the $z$-direction, the
confinement is in the $x$-direction, and the width of the transistor is found
along the $y$-direction. The source and drain regions with the lengths
$L_{\mathrm{S}}$ and $L_{\mathrm{D}}$, respectively, are $n$-doped by
phosphorus with concentration $N_{\mathrm{D}}$, whereas the $p$-doped channel
with the length $L_{\mathrm{CH}}$ has concentration $N_{\mathrm{A}}$ of boron
acceptors. The lateral surface of the semiconductor brick in the channel
region is covered by the SiO$_{2}$ oxide layer, while the aluminum gate
overlays the oxide layer.

In the semiconductor, the electron motion is determined by the following
Hamiltonian~:
\begin{align}
\hat{H}  &  =\sum_{j}{\hat{H}}_{j}\nonumber\\
{\hat{H}}_{j}  &  =-\frac{\hbar^{2}}{2m_{j}^{\text{\textrm{xx}}}}%
\frac{\partial^{2}}{\partial x^{2}}-\frac{\hbar^{2}}{2m_{j}^{\text{\textrm{yy}%
}}}\frac{\partial^{2}}{\partial y^{2}}-\frac{\hbar^{2}}{2m_{j}%
^{\text{\textrm{zz}}}}\frac{\partial^{2}}{\partial z^{2}}+V(\vec{r})\;,
\label{Hame}%
\end{align}
where $\vec{r}=(x,y,z)$; $V(\vec{r})=V_{\mathrm{B}}(\vec{r})+V_{\mathrm{e}%
}(\vec{r})$ is the potential energy associated with the energy barrier and the
electrostatic field; $m_{j}^{\text{\textrm{xx}}}$, $m_{j}^{\text{\textrm{yy}}%
}$ and $m_{j}^{\text{\textrm{zz}}}$ are the effective masses of the motion
along the $x,$ $y$ and $z$-axes, respectively, of an electron of the $j$-th valley.

The Schr\"{o}dinger equation is solved within an effective mass approximation.
It is assumed that Si/SiO$_{2}$ interface is parallel to the (100) plane. The
conduction band in bulk silicon can be represented by six equivalent
ellipsoids. When an electric field is applied in the [100]~-direction, these
six equivalent minima split into two sets of subbands \cite{Ando}. The first
set of subbands (unprimed) is two fold degenerate and represents those
ellipsoids that respond with a heavy effective mass in the direction of the
applied electric field. Because of the heavier mass, the `unprimed' subbands
have relatively lower bound-state energies, as compared to the `primed'
subbands and are therefore primarily occupied by electrons \cite{Lunds}.

The electrostatic potential energy $V_{\mathrm{e}}(\vec{r})$ satisfies
Poisson's equation
\begin{equation}
\boldsymbol{\nabla \, \cdot}\left[  \varepsilon_{i}%
\boldsymbol{\nabla}V_{\mathrm{e}}(\vec{r})\right]  =\frac{e^{2}}%
{\varepsilon_{0}}\biggl(-n(\vec{r})+N_{\mathrm{D}}(\vec{r})-N_{\mathrm{A}%
}(\vec{r})\biggr)\;,\quad i=1,2\;, \label{Poisson}%
\end{equation}
where $\varepsilon_{1}$ and $\varepsilon_{2}$ are the dielectric constants of
the semiconductor and oxide layers, respectively; $n(\vec{r})$, $N_{\mathrm{D}%
}(\vec{r})$and $N_{\mathrm{A}}(\vec{r})$ are the concentrations of electrons,
donors and acceptors, respectively. The barrier potential is non-zero in both
oxide layers, and its value is denoted by a constant~$V_{\mathrm{B}}$. In our
calculations, we assume that the source electrode is grounded whereas the
potentials at the drain and gate electrodes equal $V_{\mathrm{D}}$ and
$V_{\mathrm{G}}$ respectively. The system is assumed to be uniform in the
$y$-direction. Consequently, the electron density is constant along the
$y$-axis and the edge effects in that direction are supposed to be negligible.

The study of the charge distribution in the cylindrical MOSFET with closed
gate electrode in the state of the thermodynamical equilibrium (see
Ref.~\cite{Pok}) has shown that the concentration of holes is much lower than
that of electrons, so that electron transport is found to provide the main
contribution to the current flowing through the SOI MOSFET. For that reason,
holes are neglected in the present transport calculations.

\section{The quantum Liouville equation}

In order to study the device architecture starting from the Wigner function
formalism, an equation is needed that describes the response of the Wigner
function to changing external conditions. The time evolution of the Wigner
function is derived from the quantum Liouville equation by applying the
Wigner-Weyl transformation. Neglecting inter-valley transitions in the
conduction band, the one-electron density matrix can be written as
\begin{equation}
\rho(\vec{r},\vec{r}^{\prime})=\sum_{j}\rho_{j}(\vec{r},\vec{r}^{\prime})\;,
\end{equation}
where $\rho_{j}(\vec{r},\vec{r}^{\prime})$ is the density matrix of electrons
residing in the $j$-th valley satisfying the quantum Liouville equation
\begin{equation}
\hbar\frac{\partial{\hat{\rho}}_{j}}{\partial t}=\left[  {\hat{H}}_{j}%
,{\hat{\rho}}_{j}\right]  \;. \label{liov1}%
\end{equation}
In order to impose the boundary conditions for the density matrix at the
electrodes, it is convenient to describe the quantum transport along the
$z$-axis in a phase-space representation. For that purpose, we expand the
$j$-valley density matrix into a series with respect to a complete set of
orthonormal functions $\Psi_{jps}(x,y,z)$ and rewrite the density matrix
$\rho_{j}$ in terms of $\zeta=(z+z^{\prime})/2$ and $\eta=z-z^{\prime}$
coordinates as
\begin{equation}
\rho_{j}(\vec{r},\vec{r}^{\prime})=\sum_{psp^{\prime}s^{\prime}}\frac{1}{2\pi
}\int_{-\infty}^{+\infty}\!\text{\textrm{d}}k\,\exp\left(  \mathrm{i}%
k\eta\right)  f_{jpsp^{\prime}s^{\prime}}(\zeta,k)\Psi_{jps}(x,y,z)\Psi
_{jp^{\prime}s^{\prime}}^{\ast}(x^{\prime},y^{\prime},z^{\prime})\;.
\label{ser1}%
\end{equation}
According to the symmetry of the system, these functions take the following
form~:
\begin{equation}
\Psi_{jps}(x,y,z)=\frac{1}{\sqrt{L_{\mathrm{y}}}}\,\psi_{js}(x,z)\,\exp\left(
\mathrm{i}py\right)  \;.
\end{equation}
The functions $\psi_{js}(x,z)$ are chosen to satisfy the equation
\begin{equation}
-\frac{\hbar^{2}}{2m_{j}^{\text{xx}}}\frac{\partial^{2}}{\partial x^{2}}%
\psi_{js}(x,z)+V(r,z)\psi_{js}(r,z)=E_{js}(z)\psi_{js}(x,z) \label{eqSc}%
\end{equation}
that describes the confined motion of an electron. Here $E_{js}(z)$ are the
eigenvalues of Eq.~(\ref{eqSc}) for a given value of the $z$-coordinate that
appears as a parameter. It will be shown, that $E_{js}(z)$ plays the role of
an effective potential in the channel, and that $\Psi_{jps}(x,y,z)$ is the
corresponding wavefunction of the motion along the $x$-axis at a fixed $z$.
Substituting the expansion~(\ref{ser1}) into Eq.~(\ref{liov1}), and using
Eq.~(\ref{eqSc}), we arrive at the equation for $f_{jpsp^{\prime}s^{\prime}%
}(\zeta,k)$~:
\begin{multline}
\frac{\partial f_{jpsp^{\prime}s^{\prime}}(\zeta,k)}{\partial t}=-\frac{\hbar
k}{m_{j}^{\text{zz}}}\frac{\partial}{\partial\zeta}f_{jpsp^{\prime}s^{\prime}%
}(\zeta,k)+\frac{1}{\hbar}\int\limits_{-\infty}^{+\infty}W_{jpsp^{\prime
}s^{\prime}}(\zeta,k-k^{\prime})f_{jpsp^{\prime}s^{\prime}}(\zeta,k^{\prime
})\,\mathrm{d}k^{\prime}\label{liov3}\\
\hskip-1cm-\sum_{s_{1},s_{1}^{\prime}}\int\limits_{-\infty}^{+\infty}{\hat{M}%
}_{jpsp^{\prime}s^{\prime}}^{s_{1}s_{1}^{\prime}}(\zeta,k,k^{\prime
})f_{jps_{1}p^{\prime}s_{1}^{\prime}}(\zeta,k^{\prime})\,\mathrm{d}k^{\prime
}\;,
\end{multline}
where
\begin{equation}
W_{jpsp^{\prime}s^{\prime}}(\zeta,k-k^{\prime})=\frac{1}{2\pi\mathrm{i}}%
\int\limits_{-\infty}^{+\infty}\!\mathrm{d}\eta\,\bigl(E_{jps}(\zeta
+\eta/2)-E_{jp^{\prime}s^{\prime}}(\zeta-\eta/2)\bigr)\,\exp\left(
\mathrm{i}(k^{\prime}-k)\eta\right)  , \label{w1}%
\end{equation}%
\begin{equation}
{\hat{M}}_{jpsp^{\prime}s^{\prime}}^{s_{1}s_{1}^{\prime}}(\zeta,k,k^{\prime
})=\frac{1}{2\pi}\int\limits_{-\infty}^{+\infty}\left[  \delta_{s^{\prime
}s_{1}^{\prime}}{\hat{\Gamma}}_{pss_{1}}(\zeta+\eta/2,k^{\prime}%
)+\delta_{ss_{1}}{\hat{\Gamma}}_{ps^{\prime}s_{1}^{\prime}}^{\ast}(\zeta
-\eta/2,k^{\prime})\right]  \exp\left(  \mathrm{i}(k^{\prime}-k)\eta\right)
\mathrm{d}\eta,
\end{equation}%
\begin{equation}
{\hat{\Gamma}}_{pss_{1}}(\zeta+\eta/2,k^{\prime})=\frac{\hbar}{2m_{j}%
^{\text{zz}}\mathrm{i}\;}b_{jss_{1}}(\zeta+\eta/2)+\frac{\hbar}{2m_{j}%
^{\text{zz}}}c_{jss_{1}}(\zeta+\eta/2)\left(  -\mathrm{i}\frac{\partial
}{\partial\zeta}+2k^{\prime}\right)  ,
\end{equation}%
\[
{\hat{\Gamma}}_{ps^{\prime}s_{1}^{\prime}}^{\ast}(\zeta-\eta/2,k^{\prime
})=\frac{\hbar}{2m_{j}^{\text{zz}}\mathrm{i}\;}b_{jss_{1}}(\zeta-\eta
/2)+\frac{\hbar}{2m_{j}^{\text{zz}}}c_{jss_{1}}(\zeta-\eta/2)\left(
\mathrm{i}\frac{\partial}{\partial\zeta}+2k^{\prime}\right)
\]
and
\begin{align}
b_{jss_{1}}(z)  &  =\int\psi_{js}^{\ast}(x,z)\frac{\partial^{2}}{\partial
z^{2}}\psi_{js_{1}}(x,z)\mathrm{d}x\;,\\
c_{jss_{1}}(z)  &  =\int\psi_{js}^{\ast}(x,z)\frac{\partial}{\partial z}%
\psi_{js_{1}}(x,z)\mathrm{d}x\;,\\
E_{jps}(z)  &  =E_{js}(z)+\frac{\hbar^{2}p^{2}}{2m_{j}^{\mathrm{yy}}}\;.
\end{align}
The first term in the right-hand side of Eq.~(\ref{liov3}) is derived from the
kinetic-energy operator of the motion along the $z$-axis. It is exactly the
same as the drift term of the Boltzmann equation. The second component plays
the same role as the force term does in the Boltzmann equation. The last term
in the right-hand side of Eq.~(\ref{liov3}) contains the kernel ${\hat{M}%
}_{jpsp^{\prime}s^{\prime}}^{s_{1}s_{1}^{\prime}}(\zeta,k,k^{\prime})$, which
mixes the functions $f_{jpsp^{\prime}s^{\prime}}$ with different indices
$s,s^{\prime}$. This term appears because $\psi_{js}(x,z)$ are not
eigenfunctions of the Hamiltonian (\ref{Hame}).

\section{Boundary conditions}

To describe the behavior of the SOI MOSFET through solving Eq.~(\ref{liov3}),
we need to specify the boundary conditions for the functions $f_{jpsp^{\prime
}s^{\prime}}(\zeta,k)$ that permit particles to enter and leave the system.
The SOI MOSFET is described as an open system. Being part of an electrical
circuit, it exchanges electrons with the circuit. For the present purposes,
the term \textquotedblright open system\textquotedblright\ is used here to
characterize a system that is connected to contacts (reservoirs of particles)
and the interaction between the system and a contact necessarily involves a
particle current through an interface between the system and contacts.

The quantum Liouville equation (in the Wigner-Weyl representation) is of the
first order with respect to the coordinate $\zeta$\ and does not contain
derivatives with respect to the momentum. The characteristics of the
derivative term are lines of constant momentum, and we need to supply one and
\emph{only} one boundary value at some point on each characteristic, because
the equation is of first order with respect to the coordinate $\zeta$. Thus,
the Wigner function is a natural representation for an open system (SOI
MOSFET). The implementation of the above described picture and the comparison
of Eq.~(\ref{ser1}) with the corresponding expansion of the density matrix in
the equilibrium state, leads to the following boundary conditions
\cite{Frensley1,Frensley,foot}:\newline%
\begin{align}
f_{jmsm^{\prime}s^{\prime}}(0,k)  &  =2\delta_{ss^{\prime}}\delta_{mm^{\prime
}}\bigl[\exp\beta(E_{jspk}-E_{\mathrm{FS}})+1\bigr]^{-1}\;,\qquad
k>0\;,\nonumber\\
f_{jmsm^{\prime}s^{\prime}}(L,k)  &  =2\delta_{ss^{\prime}}\delta_{mm^{\prime
}}\bigl[\exp\beta(E_{jspk}-E_{\mathrm{FD}})+1\bigr]^{-1}\;,\qquad k<0\;,
\label{bound}%
\end{align}
where the total energy is $E_{jspk}=\hbar^{2}k^{2}/2m_{j}^{\mathrm{zz}}%
+\hbar^{2}p^{2}/2m_{j}^{\mathrm{yy}}+E_{js}(0)$ for an electron entering from
the source electrode $(k>0)$ and $E_{jspk}=\hbar^{2}k^{2}/2m_{j}^{\mathrm{zz}%
}+\hbar^{2}p^{2}/2m_{j}^{\mathrm{yy}}+E_{js}(L)$ for an electron entering from
the drain electrode $(k<0)$. In Eq.~(\ref{bound}) $\beta=1/{k_{\mathrm{B}}T}$
is the inverse thermal energy, while $E_{\mathrm{FS}}$ and $E_{\mathrm{FD}}$
are the Fermi energy levels in the source and in the drain respectively. Note,
that Eq.~(\ref{bound}) meets the requirement of imposing only one boundary
condition on the function $f_{jpsp^{\prime}s^{\prime}}(\zeta,k)$ at a fixed
value of $k$ as Eq.~(\ref{liov3}) is a first-order differential equation with
respect to $\zeta$.

\section{Ballistic regime of the SOI MOSFET}

The functions $f_{jpsp^{\prime}s^{\prime}}(\zeta,k)$, which are introduced in
Eq.~(\ref{ser1}), are used in calculations of the current and the electron
density. The expression for the electron density follows directly from the
density matrix as $n(\vec{r})=\rho(\vec{r},\vec{r})$. In terms of the
functions $f_{jpsp^{\prime}s^{\prime}}(\zeta,k)$, the electron density can be
written as follows~:
\begin{equation}
n(\vec{r})=\frac{1}{2\pi}\sum_{jpsp^{\prime}s^{\prime}}\;\int\limits_{-\infty
}^{+\infty}f_{jpsp^{\prime}s^{\prime}}(z,k)\text{ }\mathrm{d}k\text{ }%
\Psi_{jps}(x,y,z)\Psi_{jp^{\prime}s^{\prime}}^{\ast}(x,y,z)\;. \label{nr}%
\end{equation}
The current density can be expressed in terms of the density
matrix~\cite{Landau}~:
\begin{equation}
\vec{J}(\vec{r})=\sum_{j}\left.  \frac{e\hbar}{2m_{j}\,}\left(  \frac
{\partial}{\partial\vec{r}}-\frac{\partial}{\partial\vec{r}^{\prime}}\right)
\rho_{j}(\vec{r},\vec{r}^{\prime})\right\vert _{\vec{r}=\vec{r}^{\prime}}\;.
\label{current}%
\end{equation}
The total current passing through the cross-section of the structure at a
point $z$, can be obtained by an integration over the transverse coordinates.
Substituting the expansion (\ref{ser1}) into Eq.~(\ref{current}) and
integrating over $x$ and $y$, we find
\begin{equation}
J=e\sum_{j,p,s}\frac{1}{2\pi}\int\limits_{-\infty}^{+\infty}\mathrm{d}%
k\frac{\hbar k}{m_{j}^{\mathrm{zz}}}f_{jpsps}(z,k)-\frac{2e\hbar}%
{m_{j}^{\mathrm{zz}}}\sum_{j,p,s<s^{\prime}}c_{jss^{\prime}}(z)\int
\limits_{-\infty}^{+\infty}\mathrm{d}k\ \Im\bigl(f_{jpsps^{\prime}%
}(z,k)\bigr)\;, \label{cur3}%
\end{equation}
where $\Im(f)$ is the imaginary part of $f$. The first term in the right-hand
side of Eq.~(\ref{cur3}) is similar to the expression for a current of the
classical theory~\cite{Landau}. The second term, which only depends on the
non-diagonal functions $f_{jpsps^{\prime}}$, takes into account the effects of
intermixing between different states of the transverse motion. \newline Since
${\hat{M}}_{jpsp^{\prime}s^{\prime}}^{s_{1}s_{1}^{\prime}}$ couples functions
$f_{jpsp^{\prime}s^{\prime}}(\zeta,k)$ with different quantum numbers ($jps$),
it can be interpreted as a collision operator, which describes transitions of
electrons between different quantum states of the transverse motion. As shown
in Ref. \cite{Pok}, the third term in the right-hand side of Eq.~(\ref{liov3})
is significant only in the close vicinity of the $p$--$n$ junctions.
Therefore, this term is assumed to give a small contribution to the charge and
current densities. Using the above assumption, we can treat the last term in
the right-hand side of Eq.~(\ref{liov3}) as a perturbation. Here after, we
investigate the steady state of the system in a zeroth order approximation
with respect to the operator ${\hat{M}}_{jpsp^{\prime}s^{\prime}}^{s_{1}%
s_{1}^{\prime}}$. Neglecting the latter, one finds that, due to the boundary
conditions (\ref{bound}), all non-diagonal functions $f_{jpsp^{\prime
}s^{\prime}}(\zeta,k)$ ($p\neq p^{\prime}$ or $s\neq s^{\prime}$) are equal to
zero \cite{old}. Furthermore, summation over $p$ in Eq.~(\ref{liov3}) gives
\begin{equation}
\frac{\hbar k}{m_{j}^{\mathrm{zz}}}\frac{\partial}{\partial\zeta}f_{js}%
(\zeta,k)-\frac{1}{\hbar}\int\limits_{-\infty}^{+\infty}W_{js}(\zeta
,k-k^{\prime})f_{js}(\zeta,k^{\prime})\ \mathrm{d}k^{\prime}=0 \label{liov6}%
\end{equation}
with
\begin{equation}
f_{js}(\zeta,k)=\frac{1}{2\pi}\sum_{p}f_{jpsps}(\zeta,k)\;. \label{full}%
\end{equation}
In Eq.~(\ref{liov6}) the following notation is used~:
\begin{equation}
W_{js}(\zeta,k)=-\frac{1}{2\pi}\int\limits_{-\infty}^{+\infty}\left(
E_{js}(\zeta+\eta/2)-E_{js}(\zeta-\eta/2)\right)  \sin(k\eta)\;\mathrm{d}%
\eta\;.
\end{equation}
The effective potential $E_{js}(z)$ can be interpreted as the bottom of the
subband $(j,s)$ in the channel. The function $f_{js}(\zeta,k)$ is referred to
as a partial Wigner function describing electrons, that are traveling through
the channel in the inversion layer subband~$(j,s)$.

\section{Electron scattering}

In this section, we consider the electron scattering due to acoustic phonons,
impurities and surface roughness at the Si-SiO$_{2}$ interface. For this
purpose, we introduce a Boltzmann-like single collision
term~\cite{Landau,Carrut}, which in the present case has the following form
\begin{equation}
\mathrm{St}\ f_{jsmk}=\sum_{j^{\prime}s^{\prime}p^{\prime}k^{\prime}}\left(
P_{jspk,j^{\prime}s^{\prime}p^{\prime}k^{\prime}}f_{j^{\prime}s^{\prime
}p^{\prime}k^{\prime}}-P_{jspk,j^{\prime}s^{\prime}p^{\prime}k^{\prime}%
}f_{jspk}\right)  \;. \label{s1}%
\end{equation}
As noted above, we have neglected all transitions between quantum states with
different sets of quantum numbers $j$ and $s$. In general, the density matrix
has off-diagonal elements between different subbands in the presence of
inter-subband scattering processes. But when the broadening of inverse layer
levels due to inter~-subband scattering is sufficiently small in comparison
with subband energy separations, we can neglect such off~-diagonal parts and
use the transport equation (\ref{liov6}), which becomes a set of decoupled
equations for distribution functions associated with each subband $(j,s)$,
i.e. in this case the electrons in the different subbands can be considered
independent current carriers.

The transition rates depend upon the full set of quantum numbers of each
three-dimensional state. Because the present calculations consider only the
longitudinal momentum $k$, the scattering rates must be \textquotedblright
projected\textquotedblright\ onto the one-dimensional model. To do so, we
first assume that the distribution of electrons with respect to the motion
along the $y$-direction is the normalized Maxwellian distribution at a fixed
temperature, because in the source and drain contacts the distribution of
electrons over the quantum states of the motion along the $y$-direction
corresponds to an equilibrium distribution. Consequently, due to the symmetry
of the system, it is reasonable to assume that across the whole structure the
electron distribution is given by
\begin{equation}
f_{jspk}(\zeta)=f_{js}(\zeta,k)\;w_{jsp}\;, \label{s2}%
\end{equation}
where the integration measure
\begin{equation}
w_{jsp}=\sqrt{\frac{\hbar^{2}\beta}{2m_{j}^{\mathrm{yy}}\pi}}\exp\left(
-\frac{\beta\hbar^{2}p^{2}}{2m_{j}^{\mathrm{yy}}}\right)  \;\mathrm{d}p
\label{s3}%
\end{equation}
is the normalized Maxwellian distribution function with respect to the
momentum~$p$. The integration of Eq.~(\ref{s1}) over the angular momentum
gives the collision term
\begin{equation}
\mathrm{St}\ f_{js}(\zeta,k)=\sum_{k^{\prime}}\left(  P_{js}(\zeta
,k,k^{\prime})f_{js}(\zeta,k^{\prime})-P_{js}(\zeta,k^{\prime},k)f_{js}%
(\zeta,k)\right)  \;, \label{s4}%
\end{equation}
where
\begin{equation}
P_{js}(\zeta,k,k^{\prime})=\sum_{pp^{\prime}}P_{jspk,jsp^{\prime}k^{\prime}%
}w_{jsp^{\prime}}\left(  \frac{2\pi}{L_{\mathrm{y}}}\right)  ^{2}\;.
\label{s5}%
\end{equation}
This collision term is directly incorporated into the quantum Liouville
equation (\ref{liov6}) as
\begin{equation}
{\tilde{W}}_{js}(\zeta,k,k^{\prime})=W_{js}(\zeta,k-k^{\prime})+\frac{L\hbar
}{2\pi}\left(  P_{js}(\zeta,k,k^{\prime})-\delta_{k,k^{\prime}}\sum
_{k^{\prime}}P_{js}(\zeta,k^{\prime},k)\right)  \;, \label{s6}%
\end{equation}
where ${\tilde{W}}_{js}(z,k,k^{\prime})$ is the modified force term in
Eq.~(\ref{liov6}). \newline The scattering rates in the first Born
approximation \cite{Schiff} are evaluated according to the Fermi's golden
rule
\begin{equation}
P_{jsmk,jsm^{\prime}k^{\prime}}=\frac{2\pi}{\hbar}\left\vert \left\langle
jsp^{\prime}k^{\prime}\right\vert {\hat{H}}_{\mathrm{int}}\left\vert
jspk\right\rangle \right\vert ^{2}\delta\left(  E_{jsp^{\prime}k^{\prime}%
}-E_{jspk}\right)  \;, \label{s7}%
\end{equation}
where ${\hat{H}}_{\mathrm{int}}$ is the Hamiltonian of the electron-phonon,
the electron-impurity or electron-interface roughness interaction. The
electron-electron interaction is taken into account by the Hartree potential
in the single electron Hamiltonian.

Details of the scattering mechanisms are presented in Appendix.

\section{Numerical model}

The system under consideration consists of regions with high (the source and
drain) and low (the channel) concentrations of electrons. The difference of
the electron distribution in the corresponding regions would produce a
considerable inaccuracy if we would have attempted to directly construct a
finite-difference analog of Eq.~(\ref{liov6}). It is worth mentioning that, in
the quasi-classical limit, i.e. $E_{js}(\zeta+\eta/2)-E_{js}(\zeta
-\eta/2)\approx\frac{\partial E_{js}(\zeta)}{\partial\zeta}\eta$,
Eq.~(\ref{liov6}) leads to the Boltzmann equation with an effective potential
which has the following exact solution in equilibrium ($E_{F}=E_{FS}=E_{FD}%
$):
\begin{equation}
f_{js}^{\mathrm{eq}}(\zeta,k)=\frac{1}{\pi}\sum_{m}\left[  \exp\left(
\beta\left(  \frac{\hbar^{2}k^{2}}{2m_{j}^{\mathrm{zz}}}+E_{js}(\zeta
)+\frac{\hbar^{2}p^{2}}{2m^{\mathrm{yy}}}-E_{\mathrm{F}}\right)  \right)
+1\right]  ^{-1}\;.\label{equil}%
\end{equation}
For numerical calculations it is useful to write down the partial Wigner
distribution function as $f_{js}(\zeta,k)=f_{js}^{\mathrm{eq}}(\zeta
,k)+f_{js}^{\mathrm{d}}(\zeta,k)$. Inserting this into Eq.~(\ref{liov6}), one
obtains the following equation for $f_{js}^{\mathrm{d}}(\zeta,k)$~:
\begin{equation}
\frac{\hbar k}{m_{j}^{\mathrm{zz}}}\frac{\partial}{\partial\zeta}%
f_{js}^{\mathrm{d}}(\zeta,k)-\frac{1}{\hbar}\int\limits_{-\infty}^{+\infty
}{\tilde{W}}_{js}(z,k,k^{\prime})f_{js}^{\mathrm{d}}(\zeta,k^{\prime
})\mathrm{d}k^{\prime}=B_{js}(\zeta,k)\;,\label{liov61}%
\end{equation}
where
\begin{align}
B_{js}(\zeta,k) &  =\nonumber\\
&  \hspace{-2cm}\frac{1}{2\pi}\int\limits_{-\infty}^{+\infty}\mathrm{d}%
k^{\prime}\int\limits_{-\infty}^{+\infty}\mathrm{d}\eta\left(  E_{js}%
(\zeta+\frac{\eta}{2})-E_{js}(\zeta-\frac{\eta}{2})-\frac{\partial
E_{js}(\zeta)}{\partial\zeta}\eta\right)  \sin\left[  (k-k^{\prime}%
)\eta\right]  f_{js}^{\mathrm{eq}}(\zeta,k^{\prime}).\label{z}%
\end{align}
The unknown function $f_{js}^{\mathrm{d}}(\zeta,k)$ takes values of the same
order throughout the whole system, and therefore is suitable for numerical
computations. In the present work, we have used the finite-difference model,
which is described in Ref.~\cite{Frensley,Jen}. The position variable takes
the set of discrete values $\zeta_{i}=\Delta\zeta i$ for $\{i=0,\dots
,N_{\zeta}\}$. The values of $k$ are also restricted to the discrete set
$k_{p}=(2p-N_{k}-1)\Delta k/2$ for $\{p=1,\dots,N_{k}\}$. The mesh spacing in
the $k$ space is
\begin{equation}
\Delta_{k}=\frac{\pi\hbar}{N_{k}\Delta\zeta}\;.
\end{equation}
The choice of discrete values for $k$ follows from a desire to avoid the point
$k=0$ and the need to satisfy a Fourier completeness relation. On a discrete
mesh, the first derivative $\frac{\partial f_{js}}{\partial\zeta}(\zeta
_{i},k_{p})$ is approximated by the left-hand difference for $k_{p}>0$ and the
right-hand difference for $k_{p}<0$. It was shown in Ref.~\cite{Frensley},
that such a choice of the finite-difference representation for the derivatives
leads to a stable discrete model. Projecting the equation (\ref{liov61}) onto
the finite-difference basis gives a matrix equation $\mathbf{L}\cdot
\mathbf{f}=\mathbf{b}$. In the matrix $\mathbf{L}$, only the diagonal blocks
and one upper and one lower co-diagonal blocks are nonzero~:
\begin{equation}
\mathbf{L}=\left(
\begin{array}
[c]{ccccc}%
A_{1} & -U & 0 & \ldots & 0\\
-V & A_{2} & -U & \ldots & 0\\
0 & -V & A_{3} & \ldots & 0\\
\vdots & \vdots & \vdots & \ddots & \vdots\\
0 & 0 & 0 & \ldots & A_{N_{\zeta}-1}%
\end{array}
\right)  .
\end{equation}
Here, $A_{i}$, $U$, and $V$ are $N_{k}\times N_{k}$ matrices,
\begin{multline}
\lbrack A_{i}]_{pp^{\prime}}=\delta_{pp^{\prime}}\left(  1+\frac
{2m_{j}^{\mathrm{zz}}\Delta\zeta}{\hbar^{2}(2p-N_{k}-1)\Delta k}%
\sum_{p^{\prime\prime}}\hbar P_{js}\left(  k_{p^{\prime\prime}},k_{p^{\prime}%
}\right)  \right)  \\
-\frac{2m_{j}^{\mathrm{zz}}\Delta\zeta}{\hbar^{2}(2p-N_{k}-1)\Delta k}\left(
W_{js}(\zeta_{i},k_{p}-k_{p^{\prime}})+\hbar P_{js}\left(  k_{p},k_{p^{\prime
}}\right)  \right)  ,
\end{multline}%
\begin{equation}
\lbrack U]_{pp^{\prime}}=\delta_{pp^{\prime}}\theta\left\{  \frac{N_{k}+1}%
{2}-p\right\}  ,\quad\lbrack V]_{pp^{\prime}}=\delta_{pp^{\prime}}%
\theta\left\{  p-\frac{N_{k}+1}{2}\right\}  ,
\end{equation}
and the vectors are
\begin{equation}
\lbrack f_{i}]_{p}=f_{js}(\zeta_{i},k_{p}),\quad\mathrm{and}\quad\lbrack
b_{i}]_{p}=B_{js}(\zeta_{i},k_{p}),\quad i=1,N_{\zeta-1},\quad p=1,N_{k}.
\end{equation}
A recursive algorithm is used to solve the matrix equation $\mathbf{L}%
\cdot\mathbf{f}=\mathbf{b}$. Invoking downward elimination, we are dealing
with $B_{i}=(A_{i}-VB_{i-1})^{-1}U$ and $c_{i}=(A_{i}-VB_{i-1})^{-1}%
(b_{i}+Vc_{i-1})\quad(i=1,\dots,N_{\zeta})$ as the relevant matrices and
vectors. Then, upward elimination eventually yields the solution $f_{i}%
=B_{i}f_{i+1}+c_{i}\quad(i=N_{\zeta}-1,\dots,1)$. If a particular index of a
matrix or a vector is smaller than 1 or larger than $N_{\zeta}-1$, the
corresponding term is supposed to vanish.

In the channel, the difference between effective potentials $E_{js}(\zeta)$
with different ($j,s$) is of the order of or larger than the thermal energy
$k_{\mathrm{B}}T$. Therefore, in the channel only the lowest few inversion
layer subbands must be taken into account. In the source and drain, however,
many quantum states $(j,s)$ of the motion along the $x$-axis are strongly
populated by electrons. Therefore, we should account for all of them in order
to calculate the charge distribution. Here, we can use the fact that,
according to our approximation, the current flows only through the lowest
subbands in the channel. Hence, only for these subbands the partial Wigner
function of electrons is non-equilibrium. In other subbands, electrons are
maintained in the state of equilibrium, even when a bias is applied. So, in
Eq.~(\ref{nr}) for the electron density, we can substitute functions
$f_{jpsps}(z,k)$ of higher subbands by corresponding equilibrium functions.
Formally, adding and subtracting the equilibrium functions for the lowest
subbands in Eq.~(\ref{nr}), we arrive at the following equation for the
electron density
\begin{equation}
n(\vec{r})=n_{\mathrm{eq}}(\vec{r})+\frac{1}{L_{\mathrm{y}}}\sum_{js}%
\int\limits_{-\infty}^{+\infty}\mathrm{d}k\left(  f_{js}(z,k)\left\vert
\psi_{js}(x;z)\right\vert ^{2}-f_{js}^{\mathrm{eq}}(z,k)\left\vert \psi
_{js}^{\mathrm{eq}}(x;z)\right\vert ^{2}\right)  \;, \label{nr1}%
\end{equation}
where $n_{\mathrm{eq}}(\vec{r})$ and $\psi_{js}^{\mathrm{eq}}(x;z)$ are the
electron density and the wavefunction of the radial motion in the state of
equilibrium respectively. The summation in the right-hand side of
Eq.~(\ref{nr1}) is performed only over the lowest subbands. Since the
electrostatic potential does not penetrate into the source and drain, we
assume that the equilibrium electron density \textit{in these regions} is
described to sufficient accuracy by the Thomas-Fermi approximation :
\begin{equation}
n_{\mathrm{eq}}(\vec{r})=N_{\mathrm{C}}\frac{2}{\sqrt{\pi}}F_{1/2}%
\bigl(\beta(eV(r,z)+E_{\mathrm{F}}-E_{C})\bigr), \label{nsd}%
\end{equation}
where the Fermi integral is defined by
\begin{equation}
F_{1/2}(x)=\int_{0}^{\infty}\!\frac{\sqrt{t}}{\exp(t-x)+1}\mathrm{d}\,t.
\label{fint}%
\end{equation}
Here $N_{\mathrm{C}}$ is the effective density of states in the conduction
band and $E_{\mathrm{F}}$ denotes the Fermi level of the system in equilibrium.

\section{Results and discussion}

For the numerical simulation of the SOI MOSFET we have fixed the geometrical
parameters of the system as follows. The thicknesses of the semiconductor and
oxide layers are chosen to be $t_{\operatorname{Si}}=20$~nm and
$t_{\mathrm{ox}}=2$~nm respectively; the width of the MOSFET along the
$y$-axis is taken to be 50~nm, the lengths of the drain and source are $20$~nm
each, while the barrier potential $V_{\mathrm{B}}$ at the Si/SiO$_{2}$
interface equals 3.2 eV. The values of the channel length considered for the
numerical calculations are $L_{\mathrm{CH}}=20,\,30,\,40$~nm. We further
assume that the source electrode is grounded, whereas the potentials at the
drain and gate electrodes equal $V_{\mathrm{D}}$ and $V_{\mathrm{G}}$
respectively. All numerical simulations are carried out with $N_{\zeta}=$ 240
and $N_{k}=$ 120.\ The $I$ -- $V$ characteristics are computed for a gate
voltage $V_{\mathrm{G}}=1.0$~V, while the drain voltage $V_{\mathrm{D}}$
varies from 0 to 1.0~V. The temperature of the system is $T=300$~K. The doping
level of the source and drain regions is $N_{\mathrm{D}}=10^{20}$~cm$^{-3}$,
whereas the channel is doped with an acceptor concentration $N_{\mathrm{A}%
}=10^{18}$~cm$^{-3}$. In calculations of the scattering the following values
of parameters for Si are used: $\Delta=0.2,0.4$~nm, $\Lambda=1.5$~nm
\cite{Ando,Data,Green}, $\Xi_{u}=$9.2~eV, $\rho=2.3283\cdot10^{3}$~kg/m$^{3}$,
$v_{0}=8.43\cdot10^{5}$~cm/s \cite{Hell}.

The simulation of the SOI MOSFET requires a self-consistent solution of three
equations: (i) the Poisson equation (\ref{Poisson}); (ii) the Schr\"{o}dinger
equation yielding the wavefunctions describing the transverse motion in the
$x$-direction (\ref{eqSc}); (iii) the equation obeyed by the partial Wigner
distribution function (\ref{liov6}). In the present calculations the eight
lowest subbands are taken into account.

Fig. 2 illustrates the distribution of the electrostatic potential for
$V_{\mathrm{D}}=0.5$ V and for the value of the channel length $L_{\mathrm{CH}%
}=30$ nm. Clearly, the main part of the applied gate voltage voltage falls
across the insulator while, due to the surface potential confinement, the
current mainly flows in a thin layer near the Si/SiO$_{2}$ interface. At the
$p$--$n$ junctions (source-channel and drain-channel) the electrons meet
barriers across the whole semiconductor. These barriers persist even for high
values of the applied voltage $V_{\mathrm{D}}$.

In Fig. 3 the effective potential for the lowest inversion subband ($j=1,s=1$)
is plotted as a function of $z$ for different values of the applied bias
ranging from $0$ to $0.5$~V, $V_{\mathrm{G}}=1~$V, $L_{\mathrm{CH}}=30$~nm. In
the case of ballistic transport (solid lines), the applied source-drain
voltage sharply drops near the drain-channel junction. The scattering of
electrons (dashed lines) smooths out the applied voltage, which is now varying
almost linearly along the whole channel. The potential obtained by taking into
account scattering is always higher than that of ballistic case. This can be
explained by noticing that channel scattering locally increases the electron
density and therefore enhances its capability of screening the applied gate voltage.

Fig. 4 shows the effective potential for the lowest inversion subband as a
function of $z$ for different values of the applied gate voltage
$V_{\mathrm{G}}$ ranging from $0$ to $1.0$ V, $V_{\mathrm{D}}=0.5$ V,
$L_{\mathrm{CH}}=30$ nm. The effective potential drop near the source-channel
junction increases with increase gate voltage. This effect results into the
potential barrier narrowing and lowering with the gate voltage.

Fig. 5 shows the current density versus the source-drain voltage
$V_{\mathrm{D}}$ for structures with channel lengths $L_{\mathrm{CH}}=20,~40$
nm. The solid lines represent the ballistic regime, the dashed and dotted
lines reveal the effect of scattering due to ionized impurities, acoustic
phonons and surface roughness at the Si/SiO$_{2}$ interface with an average
displacement $\Delta=0.2$~nm and $\Delta=0.4$~nm respectively. At
$V_{\mathrm{D}}=0.2$ V we observe a kink in the $I$ -- $V$ characteristic. For
$V_{\mathrm{D}}<0.2$ V the derivative $V_{\mathrm{D}}/I$\ reflects the
differential resistance of the structure. Obviously, the inclusion of the
electron scattering mechanisms enhances the resistance of the structure and
softens the kink in the $I$ -- $V$ characteristic, thereby appreciably
degrading the current in the MOSFET. The collisions play an especially
important role for low source-drain voltages. The surface roughness at the
Si/SiO$_{2}$ interface gives rise to an appreciable though not dominant
contribution to the scattering in the channel. When $V_{\mathrm{D}}$ exceeds
0.2 V, a \textquotedblleft saturation regime\textquotedblright\ is attained
although the currents do not converge towards constant values. From Fig.~5 one
may conclude that a decrease of the channel length leads to an increase of the
$I$ -- $V$\ slope in the \textquotedblleft saturation regime\textquotedblright%
. This result is a consequence of a reduction of the $p$--$n$ junction barrier
potential as the length of the channel becomes shorter.

Fig.~6 shows the $I$ -- $V$ characteristics of the SOI MOSFET as a function of
the gate voltage $V_{\mathrm{G}}$. Fig.~7 illustrates the drain current
$I_{\mathrm{D}}$\ versus the gate voltage $V_{\mathrm{G}}$\ of $30$ nm
channel-length SOI MOSFET for drain voltage $V_{\mathrm{D}}=1$ V. We can see
that the \textquotedblleft threshold voltage\textquotedblright\ of the
transistor is $\simeq$ 0.0 V. The influence of the scattering on the current
diminishes as the the gate voltage lowers. This fact is due to a reduced
electron concentration resulting into a lower amplitude of the scattering
process. The decrease of the electron concentration, together with an
increasing source-channel junction barrier, leads to the reduction of the
current through the channel. Fig.~7 also informs us that we are still far
removed from the International Technology Roadmap for Semiconductors (ITRS)
requirements \cite{itrs}. In particular, whereas the on-state current
($\simeq$ 5 A/cm) is in the range of the requirements ($\simeq$ 7.5 A/cm), the
off-state current (0.5 A/cm) is far above target ($\simeq$ 0.0015 A/cm).
Therefore, further optimization of the silicon-film thickness as well as
channel-doping engineering is needed.

In Fig.~8 the contours of the absolute value of the ground-state Wigner
function ($f_{11}$) are plotted with and without the presence of scattering
for structures with channel lengths $L_{\mathrm{CH}}=20$ nm (the first
row)$,\,30$ nm (the second row)$,\,40$~nm (the third row). In these figures
the first column corresponds to the ballistic regime, the second and the third
correspond to the regime with scattering due to ionized impurities, acoustic
phonons and surface roughness at the Si/SiO$_{2}$ interface with the average
displacement $\Delta=0.2$~nm and $\Delta=0.4$~nm respectively. Darker areas in
these plots indicate higher density of electrons. Far away from the $p$--$n$
junctions where the effective potential is varying almost linearly with $z$,
the partial Wigner function is positive definite and can be interpreted as a
probability distribution of electrons in phase space. When electrons are
accelerated in the inversion layer without scattering, their velocity
increases monotonously along the whole channel. Therefore, in the phase-space
representation, the distribution of ballistic electrons looks as a narrow jet
stream pointing from the source to the drain along the channel, while
scattering is seen to wash out the electron jet in the channel. Consequently,
the electron transport through the channel combines the elements of both
diffusive and ballistic motion. The mean-free-path less than the device length
does not mean that the device is near-ballistic.

In this paper we have studied in detail the effects of ultra-short channels in
an SOI MOSFET. To benefit from the SOI architecture, the thickness of the
channel should be reduced to 10 nm or less. The simultaneous scaling of the
channel length and thickness is currently under study.

\begin{acknowledgments}
We acknowledge discussions with E. P. Pokatilov and S. N. Balaban. This work
has been supported by the GOA BOF UA 2000, IUAP, FWO-V projects Nos.
G.0306.00, G.0274.01N and the WOG WO.025.99N (Belgium).
\end{acknowledgments}

\appendix*

\section{Scattering mechanisms}

We consider the electron scattering due to impurities, acoustic phonons and
surface roughness at the Si-SiO$_{2}$ interface. In case of the electron
scattering due to impurities \cite{Carrut,Chat} ${\hat{H}}_{\mathrm{int}}$ is
the electron-impurity potential minus its spatial average. The latter term has
already been included in energy level calculations (band bending
calculations), when solving Poisson and Wigner equations self-consistently.
This spatial average being, by definition, translation invariant in the layer
plane, has a zero matrix element between initial and final plane wave states
and can thus be dropped. The impurities are assumed to be randomly placed at
the sites $\vec{R}_{i}\equiv(\vec{\rho_{i}},z_{i})$. The Hamiltonian
describing scattering due to the impurities in the inversion layer is given
by
\begin{equation}
{\hat{H}}_{\text{\textrm{int}}}\left(  \vec{r}\right)  =\sum_{\vec{q}}%
V_{\vec{q}}\left(  z_{i}\right)  \exp\left(  \mathrm{i}\vec{q}\left(
\vec{\rho-\rho}_{i}\right)  \right)  ,
\end{equation}
where
\begin{equation}
V_{\vec{q}}\left(  z_{i}\right)  =\frac{2\pi e^{2}}{\varepsilon_{1}\left(
q+q_{s}F\left(  q\right)  \right)  }F\left(  q,z_{i}\right)  \;.
\end{equation}
Here $q_{s}$ is the Thomas-Fermi screening constant and $F\left(  q\right)  $
and $F\left(  q,z_{i}\right)  $ are form-factors, corresponding to the
screening impurity potential~\cite{Ando}. Because the screening effect in the
channel is very large, as seen in the Refs. \cite{Lunds,Pok}, we can model the
scattering potential of an ionized impurity as $U(\vec{r})=(4\pi
e^{2}R_{\mathrm{s}}^{2}/\varepsilon_{1})\delta(\vec{r})$, where $R_{\mathrm{s}%
}$ determines a cross-section for scattering by an impurity. Consequently, the
absolute value of the matrix element is
\begin{equation}
\left\vert \left\langle jsp^{\prime}k^{\prime}\right\vert U(\mathbf{r-R}%
_{i})\left\vert jspk\right\rangle \right\vert =4\pi e^{2}R_{\mathrm{s}}%
^{2}/\varepsilon_{1}\psi_{js}^{2}(x_{i},z_{i}).
\end{equation}
Averaging this over a uniform distribution of impurities results in the
following scattering rate
\begin{equation}
P_{jspk,jsp^{\prime}k^{\prime}}^{\mathrm{IM}}\left(  z\right)  =2\pi
C_{\mathrm{IM}}\int\limits_{0}^{L_{\mathrm{x}}}dx\psi_{js}^{4}(x,z)\delta
\left(  E_{jsp^{\prime}k^{\prime}}-E_{jspk}\right)  , \label{pi}%
\end{equation}
where%
\begin{equation}
C_{\mathrm{IM}}=\frac{1}{\hbar S}N_{\mathrm{A}}\left(  \frac{4\pi
e^{2}R_{\mathrm{s}}^{2}}{\varepsilon_{1}}\right)  ^{2}\;,
\end{equation}
$N_{\mathrm{A}}$ is the acceptor concentration and $S=L_{\mathrm{y}%
}L_{\mathrm{z}}$.

Lattice vibrations are an inevitable source of scattering, and we need to take
them into account when we calculate device characteristics at room
temperature. In this work we restrict ourselves to intra-valley and
intra-subband acoustic phonon scattering. In the approximation of an isotropic
elastic continuum we have
\begin{equation}
{\hat{H}}_{\mathrm{int}}^{\mathrm{e-ph}}=\Xi_{\mathrm{u}}\left[
D\boldsymbol{\nabla \, \cdot}\vec{u}(\vec{\rho},z)+\frac{\partial}{\partial
z}u_{z}(\vec{\rho},z)\right]  =\sum_{\vec{q}}\left(  \gamma_{\vec{q}}%
b_{\vec{q}}\exp(\mathrm{i}\vec{q\cdot r})+h.c.\right)  \;,
\end{equation}
with \cite{Yoder}%
\[
\left\vert \gamma_{\vec{q}}\right\vert ^{2}=\frac{\hbar\Xi_{\mathrm{u}}%
^{2}q^{2}}{\rho V\omega_{\vec{q}}}.
\]
$\Xi_{{}}$, $\rho$ and $\omega_{\vec{q}}$ respectively denote the deformation
potential, the mass density of the semiconductor and phonon frequency. The
matrix elements describing single phonon emission and absorption processes are
respectively given by
\begin{equation}
\left\langle jsp^{\prime}k^{\prime};N_{\vec{q}}+1\left\vert \widehat
{H}_{\mathrm{e-ph}}\right\vert jspk;N_{\vec{q}}\right\rangle =\gamma_{\vec{q}%
}^{\ast}\sqrt{N_{\vec{q}}+1}\left\langle jsp^{\prime}k^{\prime}\left\vert
\exp(-\mathrm{i}\vec{q\cdot r})\right\vert jspk\right\rangle \;,
\end{equation}%
\begin{equation}
\left\langle jsp^{\prime}k^{\prime};N_{\vec{q}}-1\left\vert \widehat
{H}_{\mathrm{e-ph}}\right\vert jspk;N_{\vec{q}}\right\rangle =\gamma_{\vec{q}%
}\sqrt{N_{\vec{q}}}\left\langle jsp^{\prime}k^{\prime}\left\vert
\exp(\mathrm{i}\vec{q\cdot r})\right\vert jspk\right\rangle ,
\end{equation}
where $N_{\vec{q}}$\ is a phonon occupation number. The emission (+) and
absorption (-) rates are
\begin{align}
P_{jspk,jsp^{\prime}k^{\prime}}^{+} &  =\frac{2\pi}{\hbar}\sum_{\vec{q}%
}\left\vert \gamma_{\vec{q}}\right\vert ^{2}\left(  N_{\vec{q}}+1\right)
\left\vert \left\langle jsp^{\prime}k^{\prime}\left\vert \exp(-\mathrm{i}%
\vec{q\cdot r})\right\vert jspk\right\rangle \right\vert ^{2}\delta\left(
E_{jsp^{\prime}k^{\prime}}-E_{jspk}+\hbar\omega_{\vec{q}}\right)  \;,\\
P_{jspk,jsp^{\prime}k^{\prime}}^{-} &  =\frac{2\pi}{\hbar}\sum_{\vec{q}%
}\left\vert \gamma_{\vec{q}}\right\vert ^{2}N_{\vec{q}}\left\vert \left\langle
jsp^{\prime}k^{\prime}\left\vert \exp(\mathrm{i}\vec{q\cdot r})\right\vert
jspk\right\rangle \right\vert ^{2}\delta\left(  E_{jsp^{\prime}k^{\prime}%
}-E_{jspk}-\hbar\omega_{\vec{q}}\right)  \;.
\end{align}
At room temperature both expressions are approximately the same. Indeed, for
$T=300$~K the thermal energy $k_{\mathrm{B}}T$ largely exceeds the phonon
energy $\hbar\omega_{\vec{q}}$, and therefore the acoustic deformation can be
considered elastic. For low energies, we can approximate the phonon occupation
number as $N_{\vec{q}}\approx k_{\mathrm{B}}T/{\hbar\omega_{\vec{q}}}\gg1$ and
adopt the Debye model $\omega_{\vec{q}}=v_{0}q$ where $v_{0}$ is the sound
velocity. Assuming equipartition of energy in the acoustic modes, we obtain
the scattering rate~:
\begin{equation}
P_{jspk,jsp^{\prime}k^{\prime}}^{\mathrm{PH}}\left(  z\right)  =2\pi
C_{\mathrm{PH}}\int\limits_{0}^{L_{\mathrm{x}}}\mathrm{d}x\,\psi_{js}%
^{4}(x,z)\,\delta\left(  E_{jsp^{\prime}k^{\prime}}-E_{jspk}\right)  ,
\end{equation}
with
\begin{equation}
C_{\mathrm{PH}}=\frac{1}{\hbar S}\Xi_{\mathrm{u}}^{2}\frac{k_{\mathrm{B}}%
T}{\rho v_{0}^{2}}.
\end{equation}

Ideally, the Si-SiO$_{2}$ interfaces would coincide with a planes $x=0$ and
$x=t_{\operatorname{Si}}$. The real interfaces however are known to fluctuate
randomly around these planes and, under the assumption that the boundary
between Si and Si-SiO$_{2}$ is still abrupt, the local deviations due to
surface roughness may be characterized by a quasi-continuous function
$\Delta(\vec{\rho})$. Here $\vec{\rho}=(y,z)$ represents the two-dimensional
position vector parallel with the ideal interfaces. We further assume that the
surface potential varies linearly with $\Delta$, i.e.
\begin{equation}
V\left(  x+\Delta\left(  \vec{\rho}\right)  \right)  \approx V\left(
x\right)  +\Delta\left(  \vec{\rho}\right)  \frac{\partial V\left(  x\right)
}{\partial x}.
\end{equation}
Neglecting the effect of surface roughness on the subband energies and wave
functions, we may calculate the scattering matrix element for surface
roughness to the lowest order in $\Delta$\ as follows \cite{Ando,Prange}:
\begin{equation}
\left\langle \;jspk\right\vert {\hat{H}}_{\mathrm{int}}^{\mathrm{SR}%
}\left\vert \;jsp^{\prime}k^{\prime}\right\rangle =\frac{\hbar^{2}}%
{2m_{j}^{\mathrm{xx}}}\left.  \frac{\partial\psi_{js}}{\partial x}%
\frac{\partial\psi_{js}}{\partial x}\right\vert _{x=0}\Delta_{\vec
{k-k}^{\prime}}=\Delta_{\vec{k-k}^{\prime}}\int_{0}^{L_{\mathrm{x}}}%
\mathrm{d}x\left\vert \psi_{js}\right\vert ^{2}\frac{\partial V\left(
x\right)  }{\partial x}\;,
\end{equation}
where $\vec{k}\equiv(k,p)$ and $\Delta_{\vec{k-k}^{\prime}}$ is the Fourier
transform of the $\Delta\left(  \vec{\rho}\right)  $%
\begin{equation}
\Delta\left(  \vec{\rho}\right)  =%
%TCIMACRO{\dsum \limits_{\vec{q}}}%
%BeginExpansion
{\displaystyle\sum\limits_{\vec{q}}}
%EndExpansion
\Delta_{\vec{q}}\exp\left(  \mathrm{i}\vec{q\cdot\rho}\right)  \;.
\end{equation}
In the Born approximation, only the magnitude squared of $\Delta_{\vec{k}%
-\vec{k}^{\prime}}$ (referred to as the power spectrum) is needed, and thus
the phase of $\Delta_{\vec{k}-\vec{k}^{\prime}}$ can be neglected. We propose
a Gaussian distribution to describe the spatial correlation of the surface
roughness
\begin{equation}
C(\vec{\rho-\rho}^{\prime})=\left\langle \Delta(\vec{\rho})\Delta(\vec{\rho
}^{\prime})\right\rangle =\Delta^{2}\exp\left(  -\frac{|\vec{\rho}-\vec{\rho
}^{\prime}|^{2}}{\Lambda^{2}}\right)  \;, \label{auto}%
\end{equation}
where $\Delta$ is the root-mean-square value of $\Delta(\vec{\rho})$, and
$\Lambda$, is referred to as the correlation length governing the the decay of
the auto-correlation function. In the above expression the brackets
$\left\langle \;\;\right\rangle $ denote sample averages \cite{Ando}. By
convolution, the power spectrum reduces to the Fourier transform of the
auto-corrleation function (\ref{auto}), that is given by
\begin{equation}
S(\vec{q})=\left\vert \Delta_{\vec{q}}\right\vert ^{2}=\pi\Delta^{2}%
\Lambda^{2}\exp\left(  -\vec{q}^{2}\Lambda^{2}/4\right)
\end{equation}
and where $\vec{q}$ is the scattered wave vector. \newline Consequently, the
interface roughness scattering rate takes the form~:
\begin{align}
P_{jspk,jsp^{\prime}k^{\prime}}^{\text{\textrm{SR}}}(z)  &  =\frac{2\pi^{2}%
}{\hbar S}\Delta^{2}\Lambda^{2}\exp\left(  -\frac{|\vec{k}-\vec{k}^{\prime
}|^{2}\Lambda^{2}}{4}\right)  \left[  \int_{0}^{L_{\mathrm{x}}}\!\mathrm{d}%
x\left\vert \psi_{js}(x,z)\right\vert ^{2}\frac{\partial V(x)}{\partial
x}\right]  ^{2}\nonumber\\
&  \times\delta\left(  E_{jsm^{\prime}k^{\prime}}-E_{jsmk}\right)  \;.
\end{align}
The total scattering rate
\begin{equation}
P_{jspk,jsp^{\prime}k^{\prime}}=P_{jspk,jsp^{\prime}k^{\prime}}^{\mathrm{IM}%
}+P_{jspk,jsp^{\prime}k^{\prime}}^{\mathrm{PH}}+P_{jspk,jsp^{\prime}k^{\prime
}}^{\mathrm{SR}}%
\end{equation}
is then inserted into Eq. (\ref{s5}) in order to obtain the one-dimensional
scattering rate
\begin{equation}
P_{js}^{\mathrm{IM+PH}}(z,k,k^{\prime})=\left(  m_{j}^{\mathrm{yy}}%
\frac{2L_{\mathrm{y}}^{2}\beta}{\hbar^{2}\pi}\right)  ^{\frac{1}{2}}\left(
C_{\mathrm{IM}}+C_{\mathrm{PH}}\right)  F\left(  \beta\frac{\hbar^{2}k^{2}%
}{2m_{j}^{\mathrm{zz}}}-\beta\frac{\hbar^{2}k^{\prime2}}{2m_{j}^{\mathrm{zz}}%
}\right)  \int\limits_{0}^{L_{\mathrm{x}}}\psi_{js}^{4}(x,z)\,\mathrm{d}x,
\end{equation}%
\begin{multline}
P_{js}^{\mathrm{SR}}(z,k,k^{\prime})=\frac{\pi}{\hbar S}\Delta^{2}\Lambda
^{2}\left(  m_{j}^{\mathrm{yy}}\frac{2L_{\mathrm{y}}^{2}\beta}{\hbar^{2}\pi
}\right)  ^{\frac{1}{2}}\exp\left(  -\frac{(k-k^{\prime})^{2}\Lambda^{2}}%
{4}\right)  \times\\
\Pi\left(  k^{2}-k^{\prime2}\right)  \left[  \int_{0}^{L_{\mathrm{x}}%
}\mathrm{d}x\left\vert \psi_{js}(x,z)\right\vert ^{2}\frac{\partial V\left(
x\right)  }{\partial x}\right]  ^{2}\;,
\end{multline}
where
\begin{multline}
\Pi\left(  x\right)  =\int_{0}^{\infty}\mathrm{d}p\frac{\exp\left[
-\beta\frac{\hbar^{2}p^{2}}{2\sqrt{m_{j}^{\mathrm{zz}}m_{j}^{\mathrm{yy}}}%
}\right]  }{\sqrt{p^{2}+x}}\times\\
\left[  \exp\left(  -\frac{\left(  p-\sqrt{p^{2}+x}\right)  ^{2}\Lambda^{2}%
}{4\sqrt{\frac{m_{j}^{\mathrm{zz}}}{m_{j}^{\mathrm{yy}}}}}\right)
+\exp\left(  -\frac{\left(  p+\sqrt{p^{2}+x}\right)  ^{2}\Lambda^{2}}%
{4\sqrt{\frac{m_{j}^{\mathrm{zz}}}{m_{j}^{\mathrm{yy}}}}}\right)  \right]  ,
\end{multline}%
\begin{equation}
F(x)=^{-x/2}{K}_{0}(|x|/2),
\end{equation}
and $K_{0}(x)$ is a McDonald function \cite{abram}.

~\newpage

~\vspace{2cm}

\begin{figure}
\centering \includegraphics[scale=1.0]{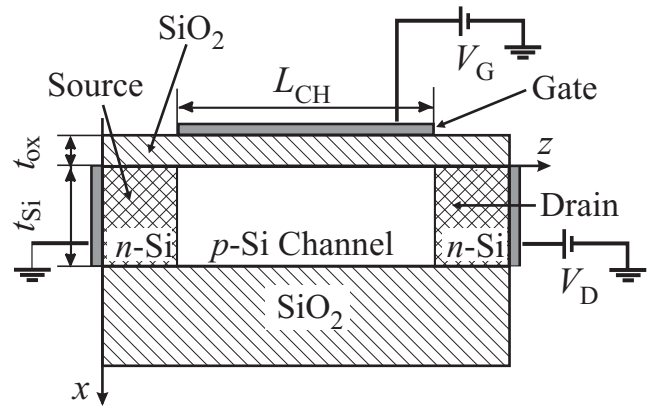}
\vspace{4cm}\caption{Scheme of the SOI MOSFET.}
\label{fig1}%
\end{figure}

~\newpage

~\vspace{2cm}

\begin{figure}
\centering \includegraphics[scale=1.0]{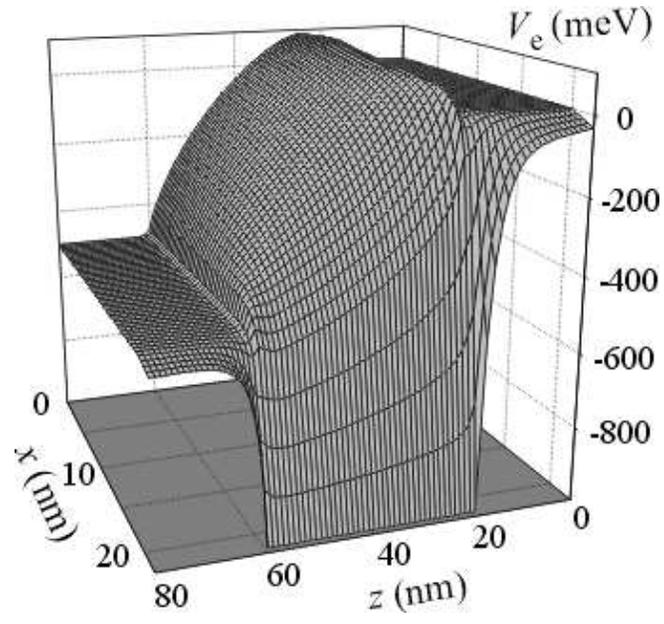}
\vspace{4cm}\caption{The electrostatic potential energy in the SOI MOSFET with
$L_{\mathrm{CH}}=40$ nm at $V_{\mathrm{G}}=1$ V and $V_{\mathrm{D}}=0.5$ V.}
\label{fig2}%
\end{figure}

~\newpage

~\vspace{2cm}

\begin{figure}
\centering \includegraphics[scale=1.0]{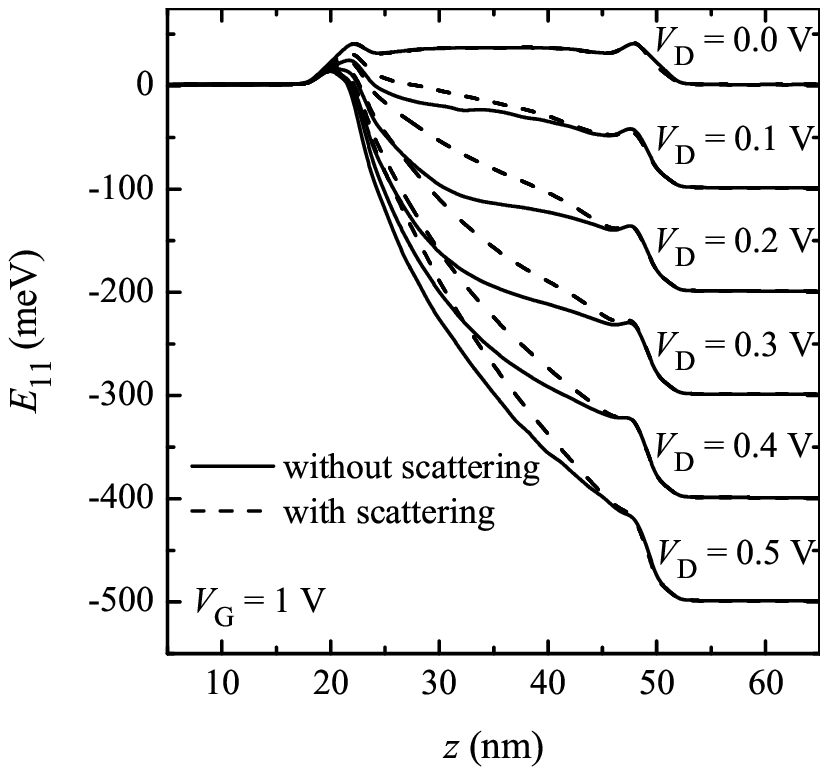}
\vspace{4cm}\caption{Effective potential as a function of $z$ for various
$V_{\mathrm{D}} $, $L_{\mathrm{CH}}=30$ nm.}
\label{fig3}%
\end{figure}

~\newpage

~\vspace{2cm}

\begin{figure}
\centering \includegraphics[scale=1.0]{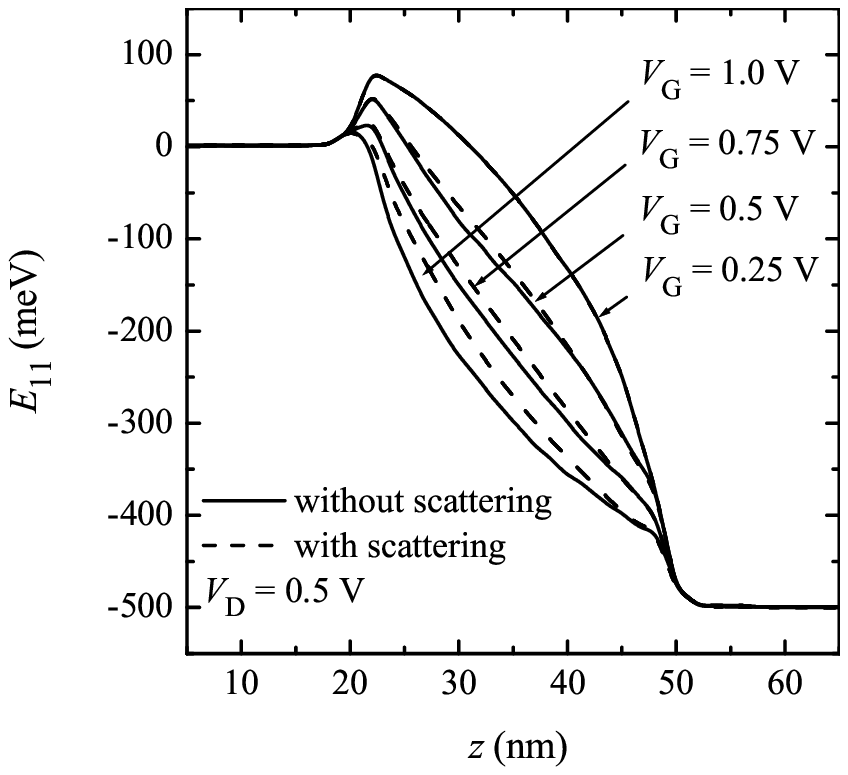}
\vspace{4cm}\caption{Effective potential as a function of $z$ for various
$V_{\mathrm{G}} $, $L_{\mathrm{CH}}=30$ nm.}
\label{fig4}%
\end{figure}

~\newpage

~\vspace{2cm}

\begin{figure}
\centering \includegraphics[scale=1.0]{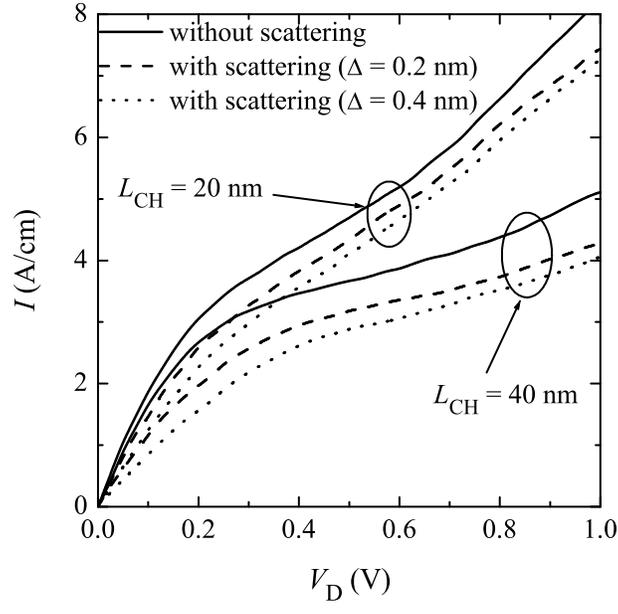}
\vspace{4cm}\caption{Current-voltage characteristics at $V_{\mathrm{G}}=1$ V for
different channel lengths.}
\label{fig5}
\end{figure}

~\newpage

~\vspace{2cm}

\begin{figure}
\centering \includegraphics[scale=1.0]{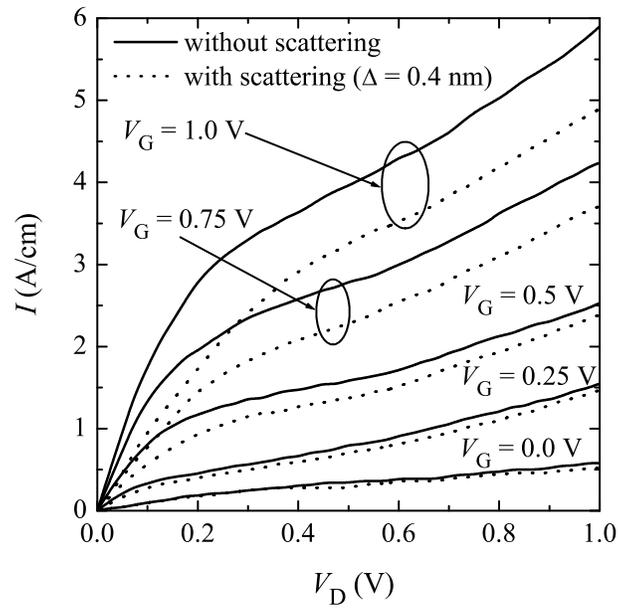}
\vspace{4cm}\caption{Current-voltage characteristics for SOI MOSFET with
$L_{\mathrm{CH} }=30$ nm.}
\label{fig6}%
\end{figure}

~\newpage

~\vspace{2cm}

\begin{figure}
\centering \includegraphics[scale=1.0]{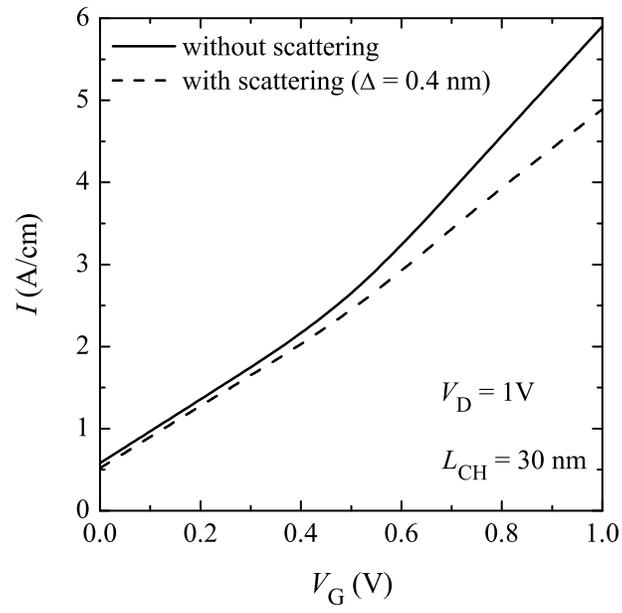}
\vspace{4cm}\caption{Drain current vs. gate voltage characteristics for drain
voltage $V_{\mathrm{D}}=1$ V.}
\label{fig7}%
\end{figure}

%~\newpage

%~\vspace{0cm}

\begin{figure}
\centering \includegraphics[scale=1.0]{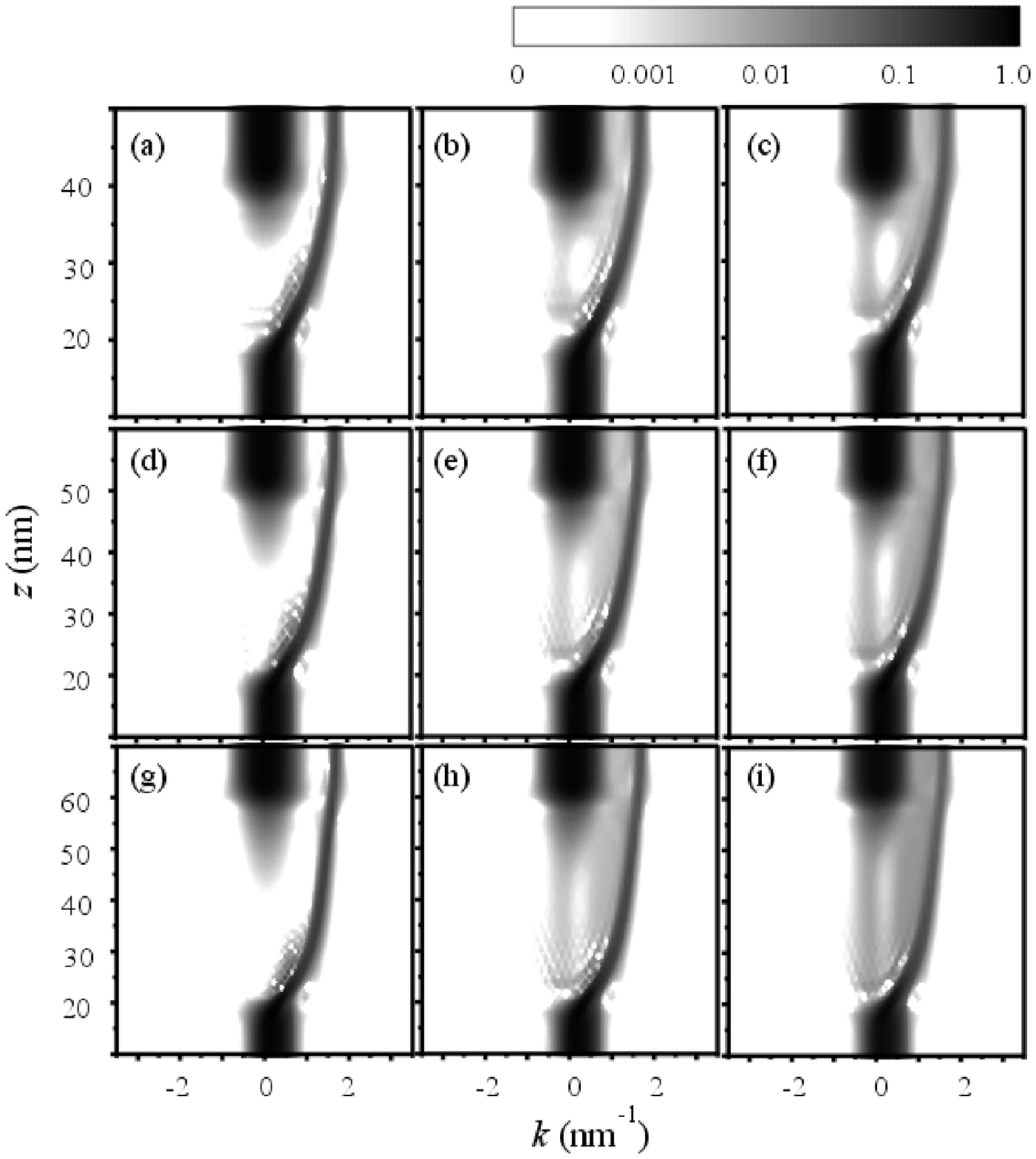}
\vspace{4cm}\caption{Contour plots of the absolute value of the partitial Wigner
distribution function $\left\vert f_{11}\left(  z,k\right)  \right\vert $ at
$V_{\mathrm{G}}=1$ V, $V_{\mathrm{D}}=0.5$ V and various channel length.
$L_{\mathrm{CH}}=20$ nm: (a) without scattering, (b) with scattering
$\Delta=0.2$ nm, (c) with scattering $\Delta=0.4$ nm; $L_{\mathrm{CH}}=30$ nm:
(d) without scattering, (e) with scattering $\Delta=0.2$ nm, (f) with
scattering $\Delta=0.4$ nm; $L_{\mathrm{CH}}=40$ nm: (g) without scattering,
(h) with scattering $\Delta=0.2$ nm, (i) with scattering $\Delta=0.4$ nm.}%
\label{fig8}
\end{figure}


\begin{thebibliography}{9999999999}
\bibitem {foot}Carriers within the active device region are injected either
from the source reservoir or the drain reservoir. Both reservoirs are assumed
in the equilibrium state, characterized by different Fermi energy levels.
\end{thebibliography}
\end{document}